\documentclass[preprint,12pt,authoryear]{elsarticle}

\usepackage{amssymb}
\usepackage{amsmath}
\usepackage{graphicx}

\journal{Icarus}

\begin{document}

\begin{frontmatter}



\title{Emissivity of Ammonia Ice}


\author{Wang, S. and Katz, J. I.}

\address{Department of Physics and McDonnell Center for the Space Sciences\\
	Washington University, St. Louis, Mo. 63130}

\begin{abstract}
Bodies in the outer Solar System may consist of, or be covered by, solid
ammonia.  Their temperature depends on its emissivity.  We
calculate that emissivity as a function of frequency, angle and polarization
and convolve it with a Planck function to obtain its integrated emissivity.

\end{abstract}


\end{frontmatter}


\section{Introduction}
\label{introduction}
The equilibrium temperature of a Solar System object is determined by the
balance between Solar heating and its infrared thermal emission.  This
depends on its albedo for Solar radiation and its infrared emissivity
averaged over a Planck function at its surface temperature.  Small bodies in
the Kuiper Belt and elsewhere in the outer Solar System, such as Ultima
Thule (486958; 2014 MU69) and elsewhere in the outer Solar System may be
made of ices.  Vapor pressures are extremely sensitive to temperature at the
temperatures $\approx 50$ K found there, so quantitative determination of
temperature is necessary to calculate the rate of evaporative loss and vapor
transport across their surfaces.  This requires quantitative knowledge of
the infrared emissivity.

Ammonia ice is of particular interest because in this temperature range its
vapor pressure varies from values so low that it is insufficient to deposit
sub-micron thick layers on colder parts of a body's surface in the age of
the Solar System to values so large that a 10 km-sized body would not
survive that time \citep{K19}.  In contrast, the vapor pressures of water
and methanol ices are so low that evaporation is negligible, while those of
methane and even carbon dioxide are so high that they would be lost entirely
from a body with negligible gravity.
\section{Calculation}
\label{calculation}
We use the complex infrared optical constants of ammonia ice \citep{MOA84}
to calculate its reflectivity $R(\nu,\theta,{\hat n})$ as a function of
frequency $\nu$, angle of incidence $\theta$ and polarization $\hat n$ from
the Fresnel relations.
The emissivity $\epsilon(\nu,\theta,{\hat n}) = 1 - R(\nu,\theta,{\hat n})$.

We assume a homogeneous half-space of solid ammonia; transmitted energy is
eventually absorbed, either by the imaginary part of the refractive index
(which is very small in most of the spectrum) or by embedded mineral matter
(dirt).  If this assumption were not made it would be necessary to specify
the depth of the ice layer, which is not known and to match electromagnetic
boundary conditions at both interfaces.  In fact, the low ($\approx 0.1$)
visible albedo of Ultima Thule \citep{S19} implies a deep homogeneous layer
of ice in which some mineral matter is embedded; if the ice were finely
divided, like snow, the visible albedo would be high because of scattering
at interfaces between ice and vacuum (or air, for terrestrial snow).

Once the mean infrared reflectivity $\langle R(T) \rangle$, averaged over a
Planck function, is known, the equilibrium temperature may be calculated.
There are two simple cases.  If sunlight is normally incident with intensity
$I_\odot$ the steady state temperature is
\begin{equation}
	\label{normal}
	T_{normal} = \left({I_\odot (1-A) \over \sigma_{SB} (1-\langle R
	\rangle)}\right)^{1/4},
\end{equation}
where $A$ is the Solar albedo, averaged over its spectrum, and $\sigma_{SB}$
is the Stefan-Boltzmann constant.  Averaging over a spherical body yields a
mean temperature
\begin{equation}
	\label{mean}
	T_{mean} = \left({I_\odot (1-A) \over 4 \sigma_{SB} (1-\langle R
	\rangle)}\right)^{1/4},
\end{equation}

The Fresnel relations for interfaces between dielectric (non-magnetic)
materials are
\begin{equation}
	R_s = \left\vert{\sqrt{1-\sin^2\theta} - \sqrt{n^2-\sin^2\theta}
	\over \sqrt{1-\sin^2\theta} + \sqrt{n^2-\sin^2\theta}}\right\vert^2
\end{equation}
and
\begin{equation}
	R_p = \left\vert{n^2\sqrt{1-\sin^2\theta} - \sqrt{n^2-\sin^2\theta}
	\over n^2\sqrt{1-\sin^2\theta} + \sqrt{n^2-\sin^2\theta}}\right\vert^2,
\end{equation}
where $s$ denotes polarization in the plane of incidence, $p$ denotes
polarization perpendicular to the plane of incidence and $n(\nu)$ is the
complex relative refractive index (the index of the solid when the wave is
incident from vacuum).

The mean reflectivity
\begin{equation}
	\langle R(T) \rangle = {1 \over 2} {\int_0^\infty\!d\nu\int_1^0\!
	d\cos\theta [R_s(\nu,\theta) + R_p(\nu,\theta)] F_\nu(T) \over
	\int_0^\infty\!d\nu\int_1^0\!d\cos\theta F_\nu(T)},
\end{equation}
where $F_\nu$ is the Planck function.  While $R_s$ and $R_p$ are properties
of the material, $R(T)$ depends on its temperature through $F_\nu(T)$.
\section{Results}
\label{results}
The results are shown in the Figures.
\begin{figure}
	\centering
	\includegraphics[width=4in]{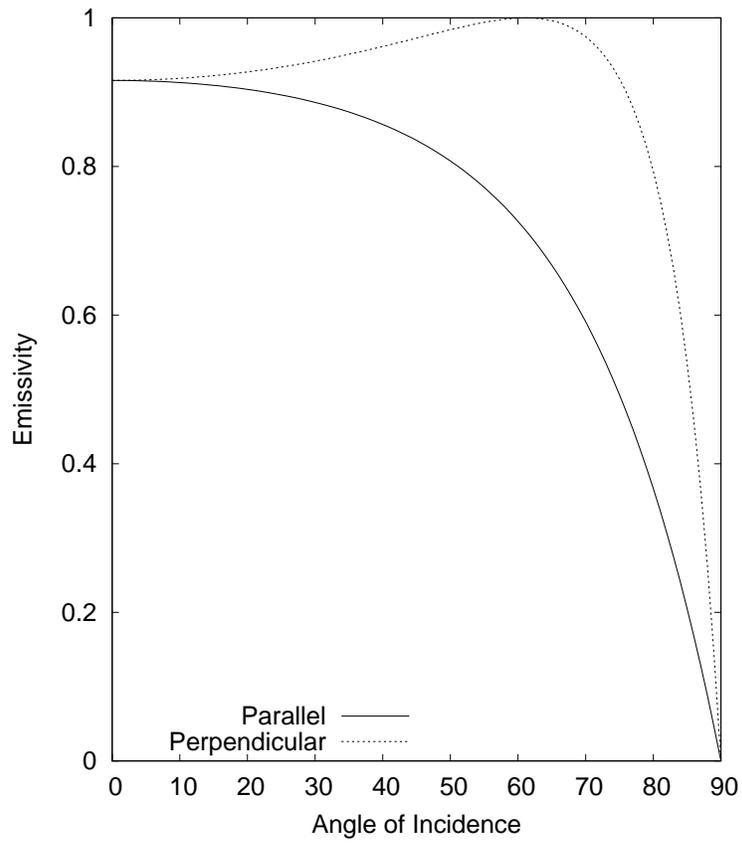}
	\caption{\label{theta} Emissivity ($1-R$) for infrared radiation
	with $\lambda = 50\mu$ as a function of angle of incidence.}
\end{figure}
\begin{figure}
	\centering
	\includegraphics[width=4in]{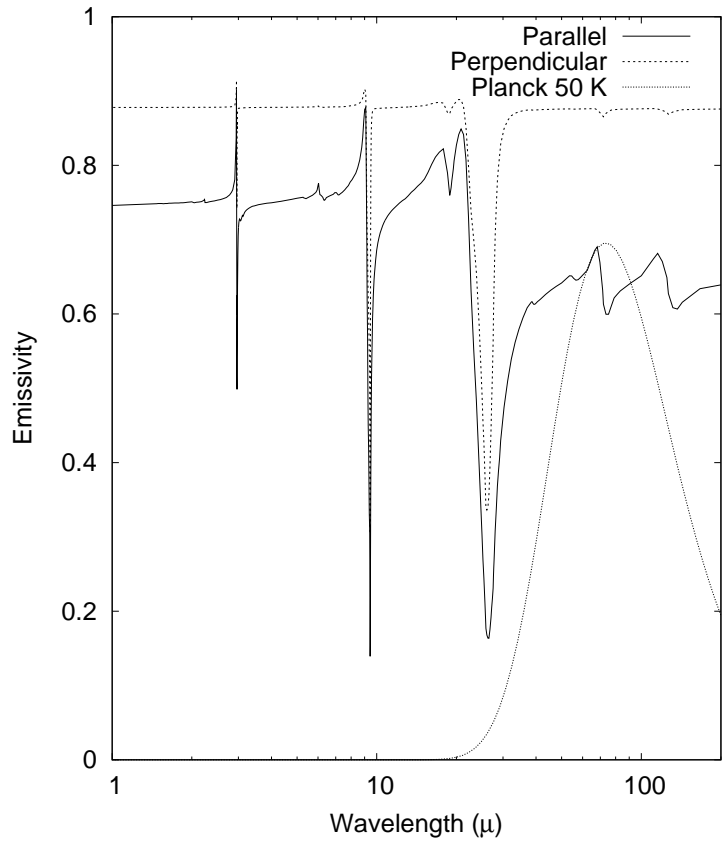}
	\caption{\label{nu} Emissivity ($1-R$) as a function of wavelength
	and polarization, averaged over solid angles.  The Planck function
	$F_\lambda$ at 50 K, with arbitrary normalization, is shown for
	comparison.}
\end{figure}
\begin{figure}
	\centering
	\includegraphics[width=4in]{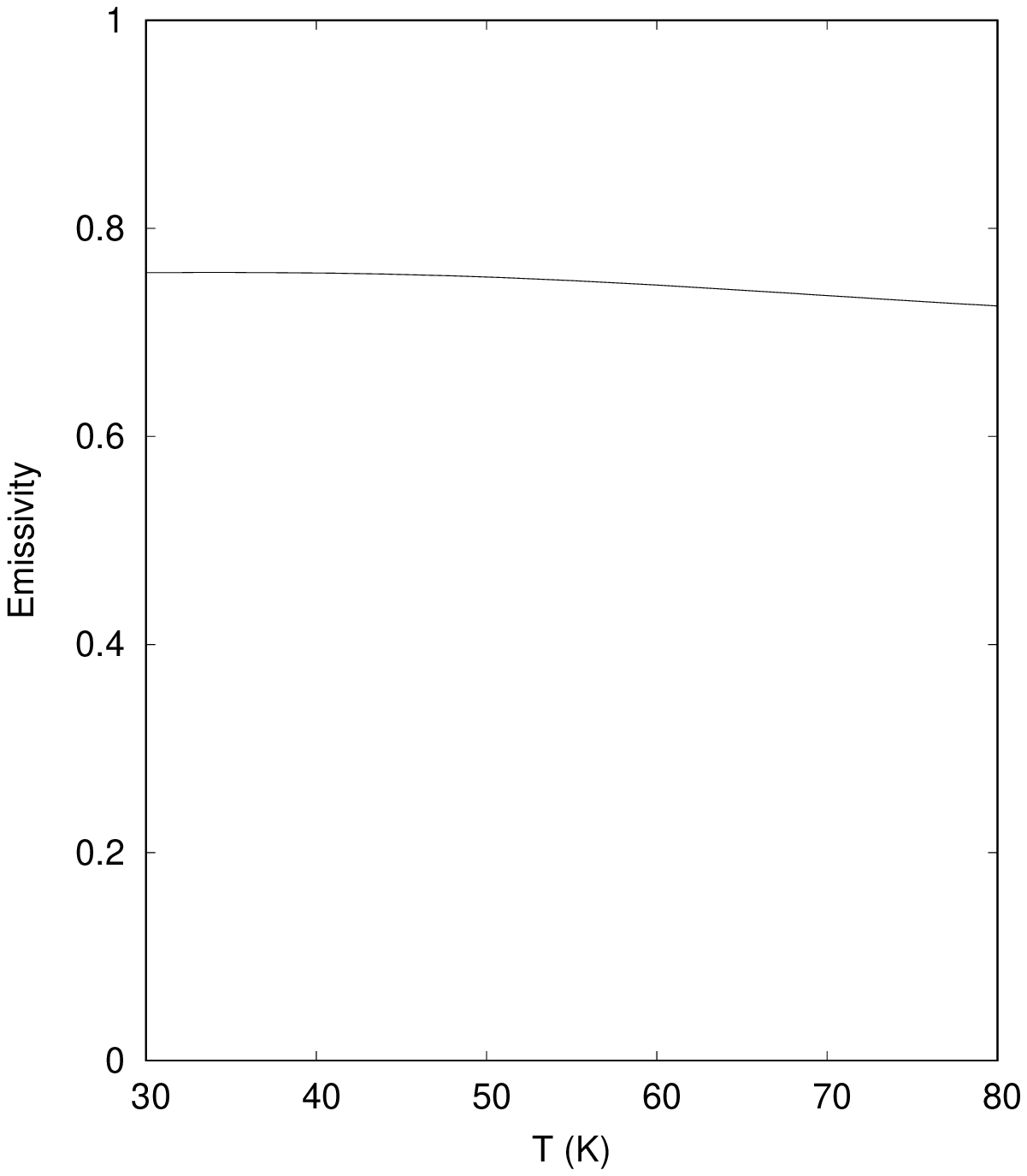}
	\caption{\label{T} Frequency-integrated emissivity ($1-\langle R
	\rangle$) as a function of temperature, averaged over solid angles,
	polarization and Planck function.}
\end{figure}
\section{Discussion}
\label{discussion}
The Planck-averaged emissivity at temperatures of interest for the Kuiper
Belt, where solid ammonia is likely to be encountered, is $\approx 0.75$ and
varies only slightly with temperature.  Inclusion of this factor in
Eqs.~\ref{normal} and \ref{mean} yields a temperature about 7.5\% higher
than would be estimated if the body were a black body radiator.  The vapor
pressure is so sensitive to temperature that this can be significant.

Finely divided ammonia ``snow'' has higher reflectivity (lower emissivity)
because light scatters at every interface between solid and vacuum.  This is
analogous to the high albedo of terrestrial water-snow.  However, it is less
extreme, because at wavelengths $\lambda \sim 100\mu$ near the 50 K black
body peak the imaginary part of the refractive index $n_i \sim 0.1$ but
varies rapidly with $\lambda$ \citep{MOA84}, suggesting $1-R \sim 0.1$.
Multiple scatterings further reduce $R$, but a quantitative calculation
would require detailed knowledge of the geometry.  However, this factor
is offset by the fact that the Solar albedo $A$ of pure ammonia snow is
likely to be close to unity (\citet{MOA84} give $n_i = 2\text{--}4 \times
10^{-5}$ for blue and red light).  Qualitatively, pure
ammonia snow will be significantly cooler than solid ammonia (with or
without mineral contamination), so that any vapor-deposited ice (hoarfrost)
will accumulate further material if the evaporation rate is significant.





\section*{Competing Interests}
The authors declare no competing interests.

\end{document}